\documentclass[reprint, superscriptaddress, amsmath,amssymb, aps, prresearch, longbibliography, floatfix]{revtex4-2}

\usepackage{amsmath}
\usepackage{graphicx}
\usepackage{ulem}
\usepackage{bm}
\usepackage{natbib}
\usepackage{dsfont}
\usepackage{tikz}
\usepackage{epsfig}
\usepackage{feynmf}
\usepackage{blindtext, rotating}
\usepackage{mathtools}
\usepackage{dsfont}
\usepackage{subcaption}
\usepackage{physics}
\usepackage{amsfonts}
\usepackage{xcolor}
\usepackage{ragged2e}
\usepackage{siunitx}
\usepackage{comment}
\usepackage{soul}
\usepackage{lipsum}
\usepackage{hyperref} 
\hypersetup{breaklinks=true, colorlinks=true, citecolor=blue, linkcolor=cyan, urlcolor=blue,filecolor=blue}

\usepackage{xcolor} 

\DeclareCaptionJustification{justified}{\justifying}

\DeclareSIUnit{\rad}{rad}

\definecolor{bright_blue}{HTML}{85C1E9}
\definecolor{middle_blue}{HTML}{2E86C1}
\definecolor{dark_blue}{HTML}{1B4F72}

\begin{document}

\title{Perturbative nonlinear feedback forces for optical levitation experiments}




\author{Oscar Kremer}
\affiliation{Center for Telecommunications Studies, Pontifical Catholic University of Rio de Janeiro, 22451-900 Rio de Janeiro, RJ, Brazil}

\author{Daniel Tandeitnik}
\affiliation{Department of Physics, Pontifical Catholic University of Rio de Janeiro, Rio de Janeiro 22451-900, Brazil}

\author{Rafael Mufato}
\affiliation{Department of Physics, Pontifical Catholic University of Rio de Janeiro, Rio de Janeiro 22451-900, Brazil}

\author{Igor Califrer}
\affiliation{Department of Physics, Pontifical Catholic University of Rio de Janeiro, Rio de Janeiro 22451-900, Brazil}

\author{Breno Calderoni}
\affiliation{Department of Physics, Pontifical Catholic University of Rio de Janeiro, Rio de Janeiro 22451-900, Brazil}


\author{Felipe Calliari}
\affiliation{Center for Telecommunications Studies, Pontifical Catholic University of Rio de Janeiro, 22451-900 Rio de Janeiro, RJ, Brazil}

\author{Bruno Melo}
\affiliation{Nanophotonic Systems Laboratory, Department of Mechanical and Process Engineering, ETH Zürich, 8092 Zürich, Switzerland}

\author{Guilherme Tempor\~ao}
\affiliation{Center for Telecommunications Studies, Pontifical Catholic University of Rio de Janeiro, 22451-900 Rio de Janeiro, RJ, Brazil}

\author{Thiago Guerreiro}
\email{barbosa@puc-rio.br}
\affiliation{Department of Physics, Pontifical Catholic University of Rio de Janeiro, Rio de Janeiro 22451-900, Brazil}

\begin{abstract}
Feedback control can be used to generate well-determined nonlinear effective potentials in an optical trap, a goal whose applications may range from non-equilibrium thermodynamics to the generation of non-Gaussian states of mechanical motion. Here, we investigate the action of an effective feedback-generated quartic potential on a levitated nanoparticle within the perturbation regime. The effects of feedback delay are discussed and predictions from the perturbation theory of a Brownian particle subjected to a quartic anharmonicity are experimentally verified.
\end{abstract}


\maketitle

\section{Introduction}

Optical levitation of nanoparticles provides a robust setup for both fundamental and applied physics \cite{millen2020optomechanics, gonzalez2021levitodynamics}, from classical stochastic thermodynamics \cite{gieseler2018levitated, gieseler2014dynamic, svak2021stochastic, sheng2023nonequilibrium} to mesoscopic quantum science \cite{romero2011large, gasbarri2021testing, pettit2019optical}. In the typical levitated optomechanics experiment, a dielectric particle is trapped in a tightly focused Gaussian beam providing, to leading order approximation, a confining harmonic potential \cite{jones2015optical, gieseler2021optical}. The particle undergoes Brownian motion due to interaction with its surrounding medium and measurements of its position correlation functions, notably the auto-correlation and the associated power spectrum, allows for the characterization of the trap's parameters \cite{hebestreit2018calibration, gieseler2021optical}. 

While the harmonic approximation is commonly employed in optical trapping, the ability to engineer potential landscapes beyond the quadratic approximation is central to optomechanics. Nonlinear force landscapes are a valuable resource to nonequilibrium Brownian machines \cite{defaveri2018dependence, defaveri2017power}, the preparation of non-classical and non-Gaussian quantum states \cite{albarelli2016nonlinearity} and matter wave interference experiments \cite{neumeier2022fast}, to mention just a few examples.
Nonlinear potential landscapes also appear in structured light optical tweezers \cite{yang2021optical}, as in double-well landscapes \cite{rondin2017direct, ricci2017optically, zhang2018nonlinearity, ciampini2020non}, structured light beams with pattern revivals \cite{da2020pattern}, cylindrical vector beams \cite{moradi2019efficient} and dark focus traps \cite{melo2020optical, almeida2023trapping}.

In these nonlinear potential landscapes, to which we refer here as \textit{nonlinear optical tweezers}, quantitative statistical description of the stochastic particle motion is significantly more complicated as it involves nonlinear stochastic differential equations. 
To make quantitative predictions regarding the statistical correlators of the trapped particle's motion we can, however, resort to perturbation theory \cite{chow2015path}. 

A perturbative method for nonlinear optical tweezers has been developed in \cite{PhysRevA.103.013110}, wherein it is possible to compute corrections to the statistical moments of particle motion, in particular the position power spectrum. The purpose of the present work is to experimentally validate these methods. Since nonlinearities in standard Gaussian optical tweezers are typically small \cite{gieseler2013thermal, cuairan2022precision}, we turn to effective feedback potential landscapes to implement nonlinear position-dependent forces upon a levitated nanosphere. 
We implement the nonlinearity via electric feedback and characterize its effects on the particle motion. 

This paper is organized as follows. In the next section, we briefly review the perturbation theory for computing corrections to the correlation functions of a trapped particle under the influence of a nonlinear force, and generalize it to include the effect of delayed forces. Since we deal with artificial electric feedback potentials relying on measurements and processing of the trapped particle's position, they imply an inherent delay to the nonlinear force and therefore accounting for the effects of this delay is essential to validating the methods of \cite{PhysRevA.103.013110}. We then describe the experimental setup used to generate nonlinear potential landscapes through electric feedback on the particle and numerically compute the effects of delay, showing that within the range of parameters employed in our experiment they are negligible. We implement a cubic force (quartic potential) on the particle and finally verify the perturbation theory by comparing the predicted center frequency of the position power spectral density with experimental results. 
We conclude with a brief discussion on the applications of artificial nonlinear forces to levitated optomechanics experiments.

\section{Theory}\label{sec:theory}

\subsection{Formulation of the perturbation theory}\label{sec:theory_no_delay}

We model the stochastic motion of a particle in a fluid at thermal equilibrium at temperature $T_{\rm{eff}}$ and under a force field $\vec{F}(\vec{r})$ using the Langevin equation,
\begin{equation}
    \ddot{\vec{r}}(t)=-\Gamma_m \dot{\vec{r}}(t)+\vec{F}(\vec{r}(t))/m+\sqrt{C}\vec{\eta}(t), \label{langevin}
\end{equation}
where $ m $ is the particle's mass, $\Gamma_m = \Gamma/m$, $C=2\Gamma k_B T_{\rm{eff}}/m^2$ with $ \Gamma $ the drag coefficient and $\vec{\eta}(t)$ is isotropic Gaussian white noise, whose components satisfy
\begin{equation}
    \mathbb{E}[\eta_{i}(t) \eta_{j}(t')] = \delta_{ij}\delta(t-t').
\end{equation}
Concentrating in the motion along the longitudinal $z$-direction, Eq. \eqref{langevin} reduces to a one dimensional Langevin equation
\begin{equation}
    \ddot{z}(t) = -\Gamma_m \dot{z}(t) + F_z(z(t))/m + \sqrt{C}\eta(t) .
    \label{Langevin_1D}
\end{equation}
For an approximately linear trapping force perturbed by nonlinear corrections, the steady state position auto-correlation $ A(t) \equiv \mathbb{E}[z(t)z(0)]$ can be perturbatively approximated.
We next summarize the perturbation theory outlined in \cite{PhysRevA.103.013110} and used throughout this work.

Consider the force acting on the particle,
\begin{equation}
    F_z(z)=-m\omega_0^2z - G_{fb}z^3 ,
\end{equation}
where the first term accounts for an optical trap with resonance frequency $\omega_{0} $ and the second term is a small nonlinear correction, which in the experiment originates from a feedback force on the particle proportional to the \textit{feedback gain} $ G_{fb} $ times a nonlinear function of the particle's position.
We define the Green's function 
\begin{equation}
    G(t) =  \frac{\sin(\Omega\, t)}{\Omega}\; \exp(-\frac{\Gamma_m t}{2})\;  H(t),
    \label{Green's function}
\end{equation}
where $\Omega = \sqrt{\omega_0^2-\Gamma_m^2/4}$ and $H(t)$ is the Heaviside step function with $H(t)=1$ for $t>0$ and $H(t)=0$ for $t\leq 0$. We introduce the response paths $\Tilde{x}(s)$ and define the Wick sum bracket $\langle(\cdots)\rangle_0$:
\begin{equation}
    \langle z(t_1)\cdots z(t_n)\Tilde{z}(s_1)\cdots \Tilde{z}(s_m)\rangle_0 = \delta_{nm}\sum_{\sigma} \prod_{j=1}^{n} G(t_j-s_{\sigma(j)})
    \label{Wick sum}
\end{equation}
where the sum goes over all permutations $\sigma$ of indexes $\{1,\ldots,n\}$. Note that the second order correlator is given by the Green function, $\langle z(t)\Tilde{z}(s)\rangle_0 = G(t-s)$.
The perturbation theory is summarized by the expression for the position auto-correlation function,
\begin{multline}
    A(t) \equiv \mathbb{E}[z(t)z(0)] = \\
    \langle z(t)z(0) e^{\frac{C}{2}\int\Tilde{z}^2(s)ds}e^{\frac{G_{fb}}{m}\int\Tilde{z}(t')z(t')^3dt'}\rangle_0,
    \label{no delay perturbation theory}
\end{multline}
where the right-hand side is defined by expanding both exponentials inside the brackets as a power series in $C$ and in $G_{fb}/m$ and interchanging summations and integrations by applying the Wick bracket $\langle(\cdots)\rangle_0$. 

\begin{figure}[t!]
    \includegraphics{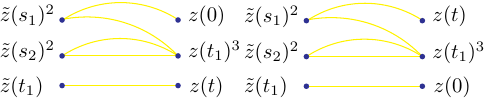}
    \caption{Diagrams for leading order correction to the position auto-correlation function of a Brownian particle subject to a non-linear optical tweezer.}
    \label{fig:diagrams}
\end{figure}

The first non-vanishing term in the expansion of Eq. \eqref{no delay perturbation theory} is
\begin{equation}
    \frac{C}{2}\int \langle z(t)z(0)\Tilde{z}(s)^2\rangle_0\,ds = C\int G(t-s)G(-s)ds \ ,
\end{equation}
which gives the auto-correlation for the case of a linear force $F_z(x)=-m\omega_0^2z $,
\begin{equation}
    A(t)_{(G_{fb}=0)}
    = \frac{C e^{-\Gamma_m|t|/2} (2\Omega \cos\Omega|t| + \Gamma_m\sin\Omega|t|)}{\Gamma_m \Omega (\Gamma_m^2+4\Omega^2)}.
    \label{Position auto-correlation for linear force}
\end{equation}
The leading order correction in the feedback gain reads,
\begin{multline}
    \Delta A(t) \equiv \\
    \frac{C^2G_{fb}}{8m}\int\langle \Tilde{z}(s_1)^2\Tilde{z}(s_2)^2\Tilde{z}(t_1) z(t_1)^3z(t)z(0)\rangle_0\,ds_1ds_2dt_1.
    \label{no delay leading order correction - formulation}
\end{multline}
Expanding the brackets using \eqref{Wick sum} would produce a sum with $5!=120$ terms, but many of these vanish since $\langle \Tilde{z}(t_1)z(t_1)\rangle = G(0)=0$. Moreover, by symmetry of the integration variables $s_1$ and $s_2$, the contribution to the integral of the non-vanishing terms is equal to the contribution of $G(t-t_1)G(-s_1)G(t_1-s_1)G(t_1-s_2)^2$ or $G(-t_1)G(t-s_1)G(t_1-s_1)G(t_1-s_2)^2$, represented by the diagrams depicted in Figure \ref{fig:diagrams}. Therefore, the integral in \eqref{no delay leading order correction - formulation} is computed by integrating these two terms over $t_1,s_1,s_2$ and multiplying both integrals by a multiplicity factor $2^3(3!)=48$. 

From the auto-correlation function perturbation $ \Delta A$ we can obtain the corrected power spectral density (PSD) of the particle motion by taking the Fourier transform. We obtain the PSD correction \cite{PhysRevA.103.013110},
\begin{equation}
    \Delta S = \frac{3G_{fb}C^{2}}{\Gamma_{m}\omega_{0}^{2}} \frac{\omega^{2} - \omega_{0}^{2}}{[\Gamma_{m}^{2}\omega^{2} + (\omega^{2} - \omega_{0}^{2})^{2}]^{2}}
\end{equation}

\noindent Note the total PSD, $ S_{(G_{fb}=0)} + \Delta S $, can be approximated to linear order in $ G_{fb} $ as,
\begin{multline}
    \frac{C}{\Gamma_m^2\omega^2 + [\omega^2-(\omega_0+\Delta \Omega)^2]^2}\approx\frac{C}{\Gamma_m^2\omega^2 + (\omega^2-\omega_0^2)^2}\\
    +4C\omega_0\Delta\Omega \frac{\omega^2-\omega_0^2}{[\Gamma_m^2\omega^2 + (\omega^2-\omega_0^2)^2]^2},
    \label{eq:expansion_shifted_oscillator}
\end{multline}
where the \textit{frequency shift} $ \Delta \Omega $ is defined by
\begin{equation}
\frac{\Delta\Omega}{2\pi} = \frac{3k_bT_{\rm{eff}}}{4 \pi m^2\omega_0^3} G_{fb} \equiv \kappa{}G_{fb}.
\label{shift_eq}
\end{equation}
We see that effectively, the nonlinear perturbation manifests as a shift in the PSD central frequency scaling linearly with the feedback gain $ G_{fb} $ and with a slope given by the constant $ \kappa $.
This is valid for small $G_{fb}$,
\begin{equation}
    G_{fb} \ll \frac{m^2\omega_0^4}{2k_bT_{\rm{eff}}}.
    \label{eq:limit}
\end{equation}
The right-hand side of \eqref{eq:limit} can be used to delimit the validity region of perturbation theory.
It is the shift $ \Delta \Omega $ in the PSD which we will use as an experimental signature of the effect of a nonlinear perturbation.

\subsection{Delayed nonlinearities}\label{sec:theory_delay}

\begin{figure}[!t]
    \includegraphics[width=0.8\linewidth]{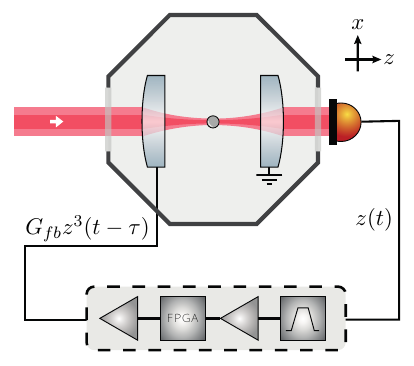}
    \caption{Experimental setup. A silica nanoparticle is trapped by an optical tweezer in vacuum. The forward scattered light is collected and sent to a photodiode, producing a signal proportional to the particle's axial coordinate, $z(t)$. An FPGA processes the signal to produce a voltage that induces a force on the trapped particle proportional to $z^3(t-\tau)$. Amplification prior to and after the FPGA enhance the maximum resolution of its analog-to-digital converter, enabling the exploration of a broader range of values for the applied electrical force.}
    \label{fig:setup}
\end{figure}

Besides nonlinear force perturbations, we will be interested in delayed forces. 
Artificially produced feedback forces will naturally be subject to electronic delay.
Accounting for the effects of such delays in perturbation theory allows us to understand the limits of validity of Eq. \eqref{no delay perturbation theory} for modelling the artificial feedback forces.
More broadly, understanding the role of delays might also enable the study of perturbative nonlinear non-Markovian stochastic dynamics \cite{innerbichler2022white}. 

We consider the generalized Langevin equation,
\begin{align}
\ddot{z}(t) = -\Gamma_m \dot{z}(t) -\omega_0^2z(t) - \frac{G_{fb}}{m}\, z(t-\tau)^3 + \sqrt{C}\,\eta(t),
\end{align}
where $\tau>0$ is a fixed (constant) time delay. The perturbation expansion for $\tau=0$ (Eq. \eqref{no delay perturbation theory}) can be generalized to
\begin{multline}
    A(t,\tau) \equiv \\
    \mathbb{E}[z(t)z(0)] = \langle z(t)z(0) e^{\frac{C}{2}\int\Tilde{z}^2(s)ds}e^{\frac{G_{fb}}{m}\int\Tilde{z}(t')z(t'-\tau)^3dt'}\rangle_0.
    \label{delayed perturbation theory}
\end{multline}
Expanding the exponentials in power series and using the Wick sum as defined in \eqref{Wick sum}, the leading correction to the auto-correlation function \eqref{Position auto-correlation for linear force} is given by the following integrals,
\begin{multline}
\Delta A(t,\tau) \propto \\
\int G(t-t_1)G(-s_1)G(t_1-s_1-\tau)G(t_1-s_2-\tau)^2dt_1ds_1ds_2 \\
+  \int G(-t_1)G(t-s_1)G(t_1-s_1-\tau)G(t_1-s_2-\tau)^2dt_1ds_1ds_2  \ .
\label{delayed leading correction}
\end{multline}
We note both integrals are multiplied by the constant $3G_{fb}C^2/m$, which we omit to avoid cluttering the notation. Evaluating the integrals leads to the corrected auto-correlation function to first order in the perturbation,

\begin{widetext}
\begin{align}\label{eq:ACF_delay}
    A(t,\tau) &= \frac{C e^{-\Gamma_m|t|/2} (2\Omega \cos\Omega|t| + \Gamma_m\sin\Omega|t|)}{\Gamma_m \Omega (\Gamma_m^2+4\Omega^2)} +
     \frac{3 C^{2} G_{fb} e^{-\Gamma_m |t|/2}}{64 m\Gamma_m^3\Omega^4\omega_0^6} \Bigg\{\nonumber\\
     &\,\quad e^{\Gamma_m\tau/2}[ 8\Gamma_m\Omega^4- 4\omega_0^2\Gamma_m^2 \Omega^2 (|t| - \tau)] \cos(\Omega (|t| - \tau)) \nonumber\\ &+ e^{\Gamma_m\tau/2}[8\Gamma_m\Omega^3\omega_0^2 (|t|-\tau)+8\Omega^5 + 4\Gamma_m^2\omega_0^2\Omega+6 \Gamma_m^2\Omega^3 ]\sin(\Omega (|t|  - \tau))\nonumber\\&+e^{-\Gamma_m\tau/2}[ \Omega^2(2 \Gamma_m^2\Omega - 8 \Omega^3) \sin(\Omega (|t| + \tau)) + 8\Gamma_m \Omega^4 \cos(\Omega (|t| + \tau))]\Bigg\}+ \order{G_{fb}^2,C^3},
\end{align}
\end{widetext}
\noindent The quantity $A(0,\tau)$ can be experimentally obtained from the area under the PSD of the particle's motion, which in turn can be related to the mean occupation number of the mechanical modes. In what follows, we use these expressions to account for the effects of delay in the artificially generated nonlinear forces, and to show that perturbation theory in the absence of delay provides a good approximation to current experiments.

\section{Experiment}

\begin{figure*}[!t]
    \centering
    \includegraphics[width=\textwidth,trim={0 0.4cm 0 0},clip]{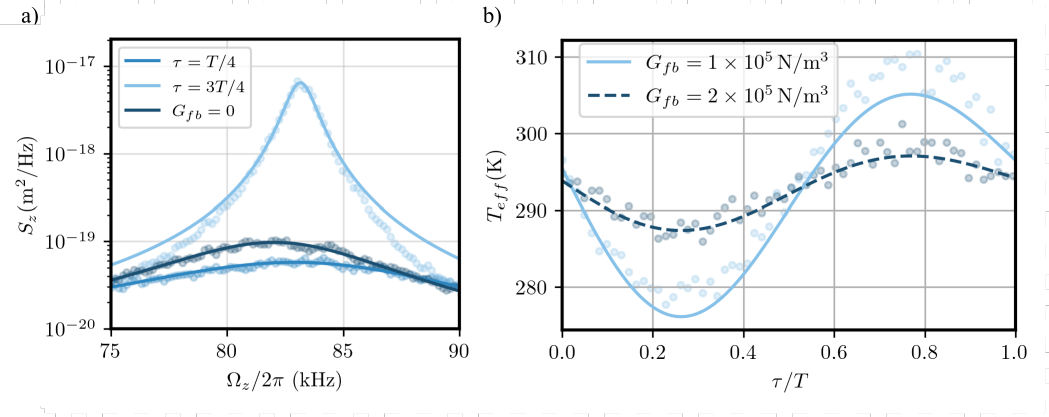}
    \caption{Effect of a delayed nonlinearity. a) Longitudinal position PSDs for the reference measurement (\protect\tikz[baseline]{\protect\draw[dark_blue, line width=0.5mm,] (0,.8ex)--++(0.5,0) ;}) in comparison to cubic feedback forces at a gain of  $G_{fb}=5.31\times10^6\,\textrm{N/m}{}^3 $ and delays of $ \tau=T/4$ (\protect\tikz[baseline]{\protect\draw[bright_blue, line width=0.5mm] (0,.8ex)--++(0.5,0) ;}) and
    $ \tau=3T/4$  (\protect\tikz[baseline]{\protect\draw[middle_blue, line width=0.5mm] (0,.8ex)--++(0.5,0);}). Here, $T$ represents the period of the particle motion along the longitudinal direction. These comparisons reveal how the introduction of a delayed cubic force can either cool or heat the particle motion. b) Numerically simulated effective temperature $ T_{\rm{eff}}$ of particle motion as a function of the delay in the cubic feedback force, displaying cooling and heating in accordance to the predictions of nonlinear delayed perturbation theory described in Sec. \ref{sec:theory_delay}. With this analysis, we conclude that the electronic delay present in our experiment, measured to be $ \tau/T =0.042\pm
 0.006$, can be safely neglected.}
    \label{fig:delay_sim}
\end{figure*}

A simplified schematic of the experimental setup is shown in Figure \ref{fig:setup}. A CW laser at \SI{780}{nm} (Toptica DL-Pro) is amplified using a tapered amplifier (Toptica BoosTa) producing up to \SI{1.5}{W} at the output of a single mode fiber, yielding a high quality Gaussian beam. The beam is expanded to overfill an aspheric lens of numerical aperture NA = 0.77 (LightPath 355330) mounted inside a vacuum chamber, which provides a tightly focused Gaussian beam to form the optical trap. A solution of silica spheres of diameter $2R = \SI{143}{nm}$ (MicroParticles GmbH) is mono-dispersed in ethanol and delivered into the optical trap using a nebulizer. Once a single particle is trapped, the pressure in the chamber is reduced to \SI{10}{mbar}. The trapped particle's axial center-of-mass (COM) motion, $z(t)$, is recorded by collecting forward scattered light with an aspheric lens of numerical aperture NA = 0.50, and directing it to a photodiode (Thorlabs PDA100A2), generating an electric signal proportional to $z(t)$. 


The signal from the detector is sent to a wide band-pass filter, amplified and then input into an FPGA. The FPGA introduces a tunable delay, raises the signal to the third power and multiplies it by a tunable gain. The output signal is then amplified once again and applied to the mount of the trapping lens, producing a voltage difference with respect to the mount of the collection lens, which is grounded. This generates an electric force at the particle position given by $G_{fb}z(t-\tau)^3$, where  $\tau$ is the total delay introduced by the electronics and $G_{fb}$ is the overall feedback gain. For more details on the generated electric field and electronics, see Appendices \ref{comsolAppendix} and \ref{electronicsAppendix}.

The electronics naturally introduce a delay to the applied position-dependent electric forces, which could lead to deviations from the predictions of the perturbation theory discussed in Sec. \ref{sec:theory_no_delay}. To qualitatively understand the effects of a delayed feedback nonlinear force, we have exaggerated the electronic delay $ \tau $ applying a cubic force 
of the form $ G_{fb}x(t-\tau)^{3} $ for $ \tau = (2\pi /  4 \omega_{0}) = T/4 $ and $ \tau = 6\pi / 4  \omega_{0} = 3T/4 $, and subsequently measured the PSDs of the particle motion along the longitudinal direction. The results can be seen in Figure \ref{fig:delay_sim}a), in comparison to the PSD of the trapped particle in the absence of nonlinear feedback. We see that depending on the delay, the particle undergoes cooling ($\tau = T/4 $) or heating ($ \tau = 3T/4 $). This can be understood as the nonlinear analogue of cold damping, where the delayed feedback signal acquires a force component proportional to the velocity \cite{tebbenjohanns2019cold, conangla2019optimal, tebbenjohanns2021quantum}. 

\begin{figure*}[htb!]
    \centering
    \includegraphics[width=\textwidth]{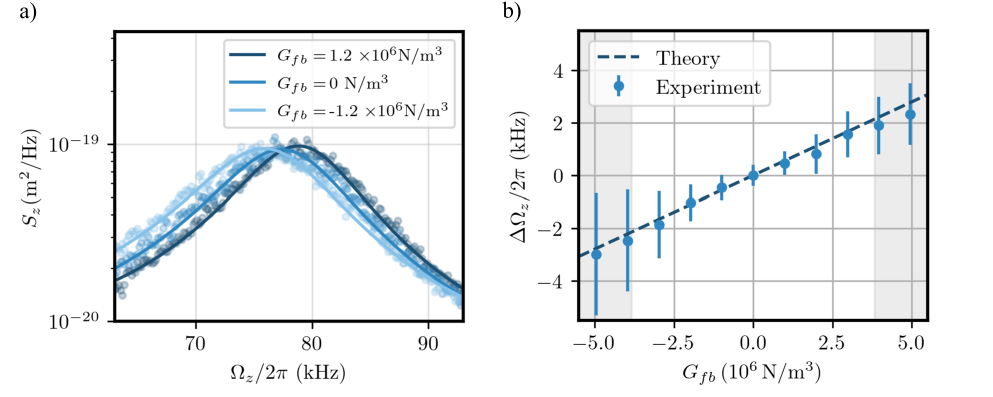}
    \caption{Verifying the predictions of perturbation theory: a) PSDs of the trapped particle's longitudinal motion under cubic force, displaying central frequency shifts. The data was taken at \SI{293}{K} and a pressure of \SI{10}{mbar}. The reference PSD (\protect\tikz[baseline]{\protect\draw[middle_blue, line width=0.5mm] (0,.8ex)--++(0.5,0);}) has a central frequency of $\SI[parse-numbers = false]{77.8}{k\hertz}$ and a shift of $\pm\SI[parse-numbers = false]{1.4}{k\hertz}$ was measured for $G_{fb}=\pm \SI[parse-numbers = false]{1.2\times 10^{6}}{\newton / \cubic\meter}$. b) Frequency shifts as a function of $G_{fb}$, verifying the prediction of perturbation theory given by Eq. \eqref{shift_eq} (dashed line). The grey shaded region marks the regime of validity for perturbation theory described in Eq. \eqref{eq:limit}. Each point corresponds to 250 seconds of data acquisition at \SI{500}{kHz} divided into 1000 traces and organized into batches of 5 traces each. All data points were collected using the same nanoparticle.
    }
    \label{fig:shift}
\end{figure*}

We can quantify the effect of delay for the case of our experiment using the theory described in Sec. \ref{sec:theory_delay}. To do that, we have simulated the particle dynamics under the influence of a delayed feedback cubic force for two different values of the feedback gain $ G_{fb} $ within the regime of perturbation theory. For each simulation, we extract the particle motion traces and compute the position variance, from which the effective temperature $T_{\rm{eff}}$ of the mechanical oscillator can be obtained. The results are plotted in Figure \ref{fig:delay_sim}b) as a function of $ \tau $, in comparison to the theoretical prediction given by Eq. \eqref{eq:ACF_delay}. The simulations confirm the qualitative cooling/heating results shown in Figure \ref{fig:delay_sim} and are in good agreement to the perturbation theory with the inclusion of delay. Notably, for the electronic delay in our experiment, characterized to be $ \tau = \SI[parse-numbers = false]{(0.518\pm
 0.074)\times 10^{-6}}{\second}$, we verify that the expected cooling/heating effects due to a delayed nonlinear feedback provide a correction to the auto-correlation at the level of $1.10 \% $ and are buried within experimental uncertainties. 
With this analysis we conclude that any effect associated to electronic delay in our experiment is negligible and the perturbation theory in the absence of delay can be used to model the effect of nonlinear perturbations.

We next proceed to verify the perturbation theory as described in Sec. \ref{sec:theory_no_delay}. We apply an effective quartic potential (cubic perturbation force) on the trapped particle generated via the position measurement feedback as described previously. PSDs of particle motion under the influence of the cubic feedback force with positive and negative feedback gains can be seen in Figure \ref{fig:shift}a). These measurements qualitatively confirm the effect of the cubic force predicted by perturbation theory as a shift in the PSD central frequency. Note the shift depends on the sign of the feedback gain, in accordance to Eq. \eqref{shift_eq}, indicating an effective hardening or softening of the optical trap due to the cubic actuation. 

To quantitatively compare the frequency shifts with the prediction from perturbation theory, we acquired the longitudinal motion PSD for different values of feedback gain $ G_{fb} $. Fitting Lorentzian functions to the PSDs we obtained the central frequency as a function of feedback gain. The result of these measurements is shown in Figure \ref{fig:shift}b), in comparison to the theoretical prediction given in Eq. \eqref{shift_eq} for our experimental parameters. Good agreement between the data and the theoretical prediction was observed within the perturbation regime, indicated by the non-shaded region of the plot. Note also that outside the regime of perturbation theory (grey shaded regions in Figure \ref{fig:shift}b)), the measured shifts fall systematically slightly bellow the predicted first order correction, consistent with the second-order correction scaling of $ \mathcal{O}(G_{fb}^{2}) $ \cite{PhysRevA.103.013110}.
Finally, the experimentally obtained angular coefficient $ \kappa_{e} $ was measured to be
\begin{align}
    \kappa_{e} &= \SI[parse-numbers = false]{(5.46\pm 0.10)\times 10^{-4}}{\hertz \cubic\meter \per\newton}
    \label{experimental_angular}
\end{align}
which compares to the theoretical prediction given the parameters for our experiment,
\begin{align}
    \kappa_{t} &= \SI[parse-numbers = false]{5.69\times 10^{-4}}{\hertz \cubic\meter \per\newton} \ .
    \label{theory_angular}
\end{align}

\section{Conclusions}

In conclusion, we have implemented a cubic nonlinear force based on position measurement feedback acting on an underdamped levitated nanoparticle. Effects of the cubic force on the particle's stochastic dynamics have been experimentally studied. In particular, shifts introduced in the particle motion power spectrum due to the presence of the cubic feedback force have been measured. We have verified that these shifts are in accordance to the predictions of the stochastic path integral perturbation theory for nonlinear optical tweezers introduced in \cite{PhysRevA.103.013110}. 
To account for the experimental imperfections due to electronic delay in the feedback, we have also extended the perturbation theory and showed that for feedback schemes currently available in levitated optomechanics experiments the effects of electronic delay can be made negligible. 

We anticipate that nonlinear electric feedback potentials will find a number of applications in levitated optomechanics experiments, both in the classical stochastic and quantum regimes. For instance, `artificial' -- i.e. feedback -- nonlinear forces could be employed in non-Gaussian state preparation protocols beyond the nonlinearities naturally present in optical potentials \cite{bateman2014near, neumeier2022fast}. Moreover, delayed nonlinear feedbacks could also be used to engineer non-conservative systems with nonlinear damping, for example of the Van der Pol type \cite{bullier2021quadratic}. In this context, the delayed perturbation theory we have introduced could be used to provide analytical predictions for feedback cooling.


\section*{Acknowledgements}
We acknowledge Bruno Suassuna for helpful discussions. T.G. acknowledges the Coordena\c{c}\~ao de Aperfei\c{c}oamento de Pessoal de N\'ivel Superior - Brasil (CAPES) - Finance Code 001, Conselho Nacional de Desenvolvimento Cient\'ifico e Tecnol\'ogico (CNPq), Funda\c{c}\~ao de Amparo \`a Pesquisa do Estado do Rio de Janeiro (FAPERJ Scholarship No. E-26/202.830/2019) and Funda\c{c}\~ao de Amparo \`a Pesquisa do Estado de São Paulo (FAPESP processo 2021/06736-5). D.T. acknowledges CAPES - Finance Code 001, and CNPq - Scholarship No. 140197/2022-2. Code and data availability: GitHub. https://github.com/QuantumAdventures/non-linearity-experiment


\bibliography{main}

\newpage

\appendix

\section{Electric field simulation}\label{comsolAppendix}

\begin{figure}[htb!] 
   \centering
   \includegraphics[width=0.95\linewidth]{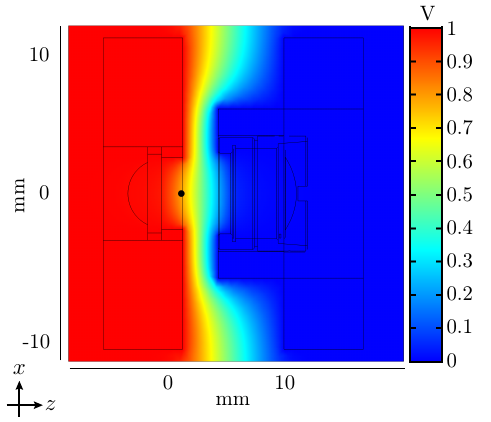}
    \caption{Electric potential generated by the electrodes' geometry for a slice in the xz plane passing through the optical axis. The contour shows the internal structure of the optical setup with the black dot marking the average position of the trapped particle, about $\SI{1.59}{mm}$ away from the flat base of the trapping lens.}\label{fig:comsolElecPot}
\end{figure}

\begin{figure*}[htb!] 
   \centering
   \includegraphics[width=\textwidth,trim={0 0.2cm 0 0},clip]{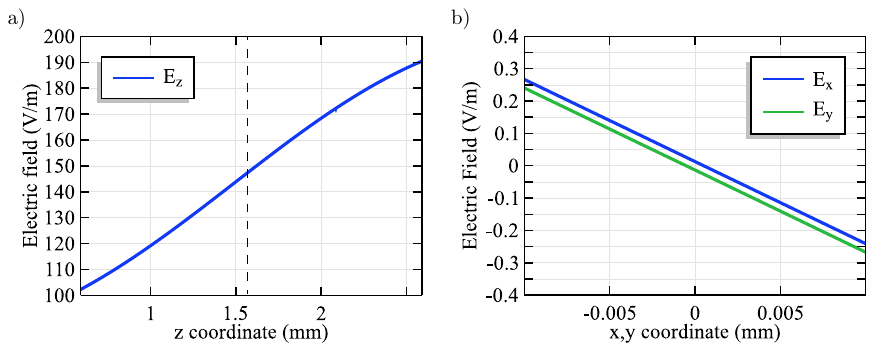}
    \caption{(a)-(b) The $z$ and $x,y$ components of the electric field in the vicinity of the trapped particle. The dashed line denotes the average position of the particle.}\label{fig:comsolElecField}
\end{figure*}

One of the experiment's central assumptions is that the electric force acting upon the trapped particle is proportional to the voltage applied to the electrodes and does not depend on its position. Moreover, due to symmetry around the optical axis, we expect the components of the electric force orthogonal to the optical axis to be negligible. To verify these assumptions, a simulation of the electric potential and electric field generated by the geometry of the optical setup was conducted using COMSOL Multiphysics software (version 5.4).

In Fig. \ref{fig:comsolElecPot}, the electrical potential between the electrodes is shown for a slice in the $xz$ plane, where the internal contour of the optical setup is displayed for clarity. The left electrode, which contains the trapping lens, is set at $\SI{1}{V}$ relative to the right one, which holds the collection lens. The black dot denotes  the average position of the trapped particle, $\SI{1.59}{mm}$ away from the flat base of the aspheric lens. Figures \ref{fig:comsolElecField}(a) and \ref{fig:comsolElecField}(b) show the electric field components in the vicinity of the particle. Considering an average amplitude of $\SI{100}{nm}$ for the COM motion, the simulation shows a percent change of roughly $0.01\%$ for the $z$ component of the electric field. Moreover, the $x$ and $y$ components are four to five orders of magnitude smaller than the $z$ component, thus providing a firm foundation for our assumptions.

\section{Electronics}\label{electronicsAppendix}

In order to apply the feedback signal, essential steps were undertaken regarding the implementation of an electronic setup aimed at preprocessing the detection signal. First, it was crucial to address a strong DC component present in the signal obtained from the photodetector. To prevent saturation of the Red Pitaya RF input used in the experiment, an analog band-pass filter was implemented for its capability to remove both DC and high-frequency components effectively. While it's common to opt for a Butterworth filter based on the Sallen-Key topology \cite{schaumann2009design}, it is important to highlight that this choice introduces an undesirable phase effect. 

\begin{figure*}[htb!] 
   \centering
   \includegraphics[width=\textwidth,trim={0 0.8cm 0 0},clip]{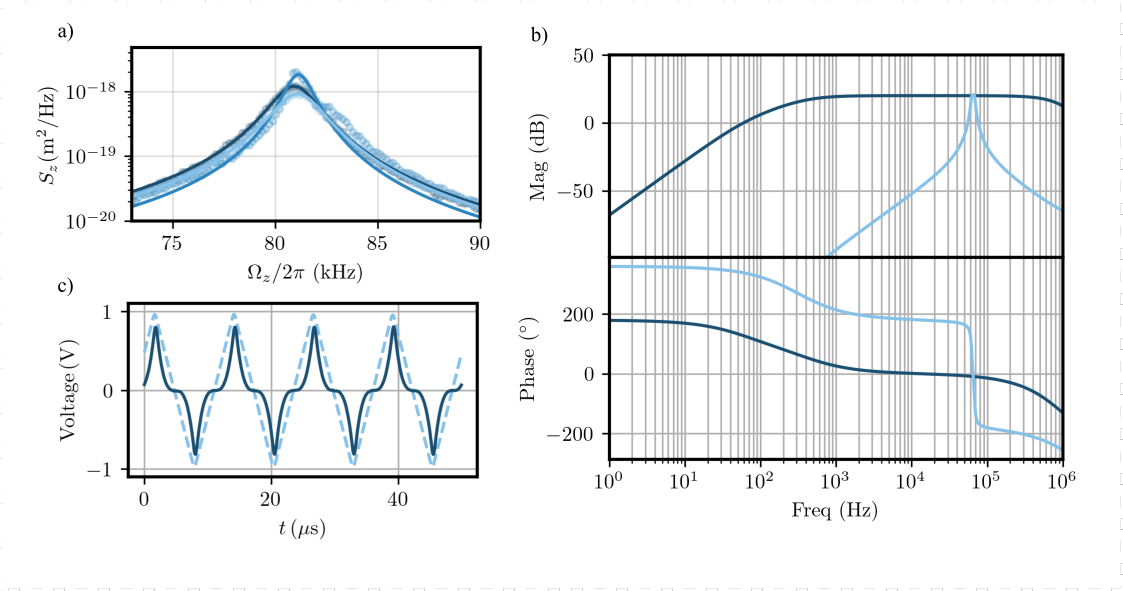}
    \caption{Filter design. (a) PSD's obtained from simulations of a tweezed nanoparticle ($\Omega_z/2\pi=\SI[parse-numbers = false]{81.5}{k\hertz}$ and $\Gamma_m=\SI[parse-numbers = false]{1.3\times 10^{4}}{\second^{-1}}$) under the influence of a cubic force. Three scenarios were considered: second-order Butterworth filter with 1 kHz bandwidth (\protect\tikz[baseline]{\protect\draw[bright_blue, line width=0.5mm] (0,.8ex)--++(0.5,0) ;}), 10 kHz bandwidth (\protect\tikz[baseline]{\protect\draw[middle_blue, line width=0.5mm] (0,.8ex)--++(0.5,0);}) and, lastly, with no filter (\protect\tikz[baseline]{\protect\draw[dark_blue, line width=0.5mm,] (0,.8ex)--++(0.5,0) ;}). (b) Bode diagrams of a highly selective Butterworth filter (\protect\tikz[baseline]{\protect\draw[bright_blue, line width=0.5mm] (0,.8ex)--++(0.5,0) ;}) and of a passive RC filter (\protect\tikz[baseline]{\protect\draw[dark_blue, line width=0.5mm,] (0,.8ex)--++(0.5,0) ;}), both circuits were simulated using LTspice XVII. (c) Results from the FPGA program. The dashed line represents the input, which is a triangular wave with a frequency of $\SI[parse-numbers=false]{81}{k\hertz}$. The solid line corresponds to the output, which is proportional to the input raised to the third power.  
    }
    \label{fig:filter-design}
\end{figure*}

As demonstrated by simulation results showed in Fig. \ref{fig:filter-design} (a), the addition of a Butterworth filter results in a shift of the PSD central frequency, which deviates from the theoretical prediction presented in \cite{PhysRevA.103.013110}. To overcome this problem a passive RC filter is used along with a non-inverter amplifier. As evident from Fig. \ref{fig:filter-design} (b), the comparison of the Bode diagrams for both topologies illustrates that the passive filter will have minimal impact on the signal phase, while simultaneously maintaining a flat band over a wider frequency range. 

The addition of a non-inverting amplifier after the band-pass filter enables the utilization of the full resolution of the ADC on the Red Pitaya board. Furthermore, a second amplifier is incorporated after the FPGA, facilitating the generation of voltage values approximately ten times higher than the board's limit. Upon characterization of both amplifiers, we found that the gains, $A_1$ and $A_2$, before and after the FPGA were measured as $\SI[parse-numbers = false]{11.00}{\volt / \volt}$ and $\SI[parse-numbers=false]{11.27}{\volt /\volt}$, respectively. These values will be necessary for the calibration of the overall feedback gain $G_{fb}$, detailed in appendix \ref{calibrationAppendix}.

In Fig. \ref{fig:filter-design} (c) we illustrate an example of input and output signals of the Red Pitaya. In order to implement the non-linear function, we employed fixed-point arithmetic—a method for representing fractional numbers within a specified range. This approach enables us to execute complex mathematical operations without suffering from information loss \cite{wilson2015design}, as is often the case with binary representation. Furthermore, it offers straightforward means of extending the code to implement higher-order polynomial functions. 
  
\section{Calibration of applied force}\label{calibrationAppendix}

To validate the theoretical predictions outlined in \cite{PhysRevA.103.013110}, it was necessary to calibrate the overall feedback gain $G_{fb}$, defined as

\begin{equation}
G_{fb} = C_{NV} A_2 A_d A_1^3 C_{mV}^3,   
\end{equation}

\noindent where $A_1$ and $A_2$ represent the gains originating from the electronic amplifiers, $A_d$ is the tunable digital gain defined within the FPGA, $C_{mV}$ is the calibration factor which converts the measured voltage into corresponding displacement in meters and  $C_{NV}$ is the transduction coefficient that establishes the connection between applied voltage across the electrodes and the resulting force applied to the particle; see appendix \ref{electronicsAppendix} for further details.

\begin{figure*}[h!] 
   \centering
   \includegraphics[width=\textwidth]{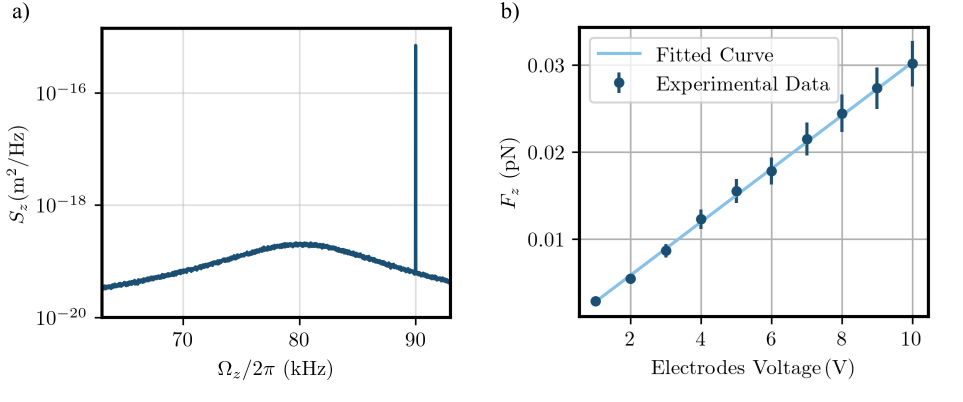}
    \caption{Electrode calibration: (a) PSD obtained from a trapped nanoparticle at 10 mbar and $T_{\rm{eff}}=\SI[parse-numbers = false]{293}{\kelvin}$ under the action of a sinusoidal drive (voltage amplitude $V_0=\SI[parse-numbers= false]{10}{\volt}$ and frequency $\omega_{dr}/2\pi = \SI[parse-numbers=false]{90}{k\hertz}$). b) Calibration curve for electrodes used to map the applied voltage to the resulting force applied on the nanoparticle. 
    }
    \label{fig:calibration_panel}
\end{figure*}

To calibrate the photodetector, 1000 traces of 0.1 seconds were collected. The PSD of the time traces is fitted by a Lorentizan distribution,

\begin{equation}
    S_V(\omega) = \frac{D}{\Gamma_m^2\omega^2+(\omega^2-\omega_0^2)^2}, 
\end{equation}
    
\noindent where $D=2\Gamma_m k_B T_{\rm{eff}}C_{mV}^2/m$; this take in consideration that $S_V(\omega)= C_{mV}^2 S_z(\omega)$ \cite{hebestreit2018calibration}. This procedure led to a calibration factor of $C_{mV}=\SI[parse-numbers = false]{(1.504\pm 0.073)\times 10^{4}}{\volt / \meter}$. After calibration of the detector, we proceed to determine the transduction coefficient, denoted as  $C_{NV}$. To obtain  $C_{NV}$, we subjected the particle to a series of sinusoidal signals with varying amplitudes and measured the particle's response in the position PSD \cite{ricci2019accurate}. For a particle subjected to Eq. \ref{Langevin_1D}, the total PSD $S_z^T(\omega)$ in the presence of an electric drive $F_{el}(t)=F_0\cos(\omega_{dr}t)$ can be expressed as \cite{ricci2019accurate},

\begin{multline}
    S_z^T(\omega) = S_z(\omega) + S_z^{el}(\omega)= \\
    \frac{2 \Gamma_m k_B T_{\rm{eff}} }{m[(\omega^2-\omega_0^2)^2+\Gamma_m^2\omega^2]}+\frac{F_0^2\tau_{el}\,\textrm{sinc}^2[(\omega-\omega_{dr})\tau_{el}]}{m^2[(\omega^2-\omega_0^2)^2+\Gamma_m^2\omega^2]},
    \label{calib_electrode_eq}
\end{multline}

\noindent with $2\tau_{el}$ being the duration of the measure. In Figure \ref{fig:calibration_panel}a), we display one of the PSDs used for the electrode calibration. The resulting calibration curve is presented in Figure \ref{fig:calibration_panel}b), which yields a transduction coefficient $C_{NV}=\SI[parse-numbers = false]{(3.06\pm 0.13)\times 10^{-15}}{\newton / \meter}$. All measurements described in the main text were performed with the same nanoparticle.

\end{document}


\title{Levitated single photon detectors -- supplementary material}

\author{Igor Califrer}
\email{califrer@puc-rio.br}
\affiliation{Department of Physics, Pontifical Catholic University of Rio de Janeiro, Rio de Janeiro 22451-900, Brazil}

\author{Tatiana Guimar\~aes}
\email{tatiana@puc-rio.br}
\affiliation{Department of Physics, Pontifical Catholic University of Rio de Janeiro, Rio de Janeiro 22451-900, Brazil}

\author{Oscar Kremer}
\email{kremer@puc-rio.br}
\affiliation{Department of Physics, Pontifical Catholic University of Rio de Janeiro, Rio de Janeiro 22451-900, Brazil}

\author{Guilherme Tempor\~ao}
\email{temporao@puc-rio.br}
\affiliation{Department of Physics, Pontifical Catholic University of Rio de Janeiro, Rio de Janeiro 22451-900, Brazil}
\author{Thiago Guerreiro}
\email{barbosa@puc-rio.br}
\affiliation{Department of Physics, Pontifical Catholic University of Rio de Janeiro, Rio de Janeiro 22451-900, Brazil}



\maketitle

\section{Kalman algorithm}

For convenience, we give a brief review of the Kalman filter algorithm adapted to our problem. We refer the reader to \cite{livro_controle} for details. 

\subsection{Kalman equations}

The Kalman filter is an algorithm for state estimation. We can think of it in terms of a black box that takes as input measurements on a dynamical system performed at a given time iteration and estimates the system's state for the next time iteration, given all prior measurement results. The algorithm works for linear dynamical systems subject to Gaussian noise.

Let the dynamical system at time $ k $ have state vector $ z_{k} $. Measurements are performed on the system, yielding results $ y_{k} $ at each time iteration. The collection of all measurement results up to time $ k $ is denoted $ \mathbb{M}_{k} $. 
The system dynamics is of the form,
\begin{eqnarray}    
z_{k} &=& f_{k-1}(z_{k-1}, u_{k-1}, w_{k-1}) \\
y_{k} &=& h_{k}(z_{k}, u_{k}, v_{k})
\end{eqnarray}
where $f_{k-1} $ is a dynamical law depending on the time parameter $ (k-1) $, the previous state of the system $ z_{k-1}$, a measured input variable $ u_{k-1}$\footnote{This is an additional known variable, which would account for noise, or feedback on the system. The important thing is that it is known, and an input of the Kalman algorithm.}, and a stochastic noise term $ w_{k-1}$; while $ h_{k} $ is a function relating the measurement result $ y_{k}$ to the system's state $ z_{k}$, the input variable $ u_{k} $ and a detector stochastic noise $ v_{k}$. Note the dynamical law only depends on the state of the system at the previous time instant: it is local in time (Markovian). 

The system is linear if the dynamical law is of the form,
\begin{eqnarray}
    f_{k-1} &=& A_{k-1} z_{k-1} + B_{k-1} u_{k-2} + w_{k-1} \\
    h_{k} &=& C_{k}z_{k} + D_{k} u_{k} + v_{k} 
\end{eqnarray}
The kalman filter is an algorithm for estimating $ z_{k}$ given all previous measurements up to time iteration $ k $, that is $ \mathbb{M}_{k} $, with $ w_{k}, v_{k} $ Gaussian-distributed random variables with a given mean and variance. 

\subsection{Estimator}

An estimator is a mathematical method for ``guessing'' the value of an unknown (random) variable given a sample of measurements. 
Define the conditional mean,
\begin{eqnarray}
    \mathbb{E}\left( z \big| \mathbb{M} \right) = \int dz \ z \ p\left( z \big| \mathbb{M} \right)
\end{eqnarray}
Several estimators can be used, for instance the Minimum Mean Error estimator,
\begin{eqnarray}
    \hat{z} = \mathrm{arg} \min_{\hat{z}}  \mathbb{E}\left( \| z - \hat{z} \| \  \big|  \ \mathbb{M} \right)
\end{eqnarray}
or the Minimum Mean Squared Error,
\begin{eqnarray}
    \hat{z} = \mathrm{arg} \min_{\hat{z}}  \mathbb{E}\left( \| z - \hat{z} \|^{2} \  \big|  \ \mathbb{M} \right)
\end{eqnarray}
In general, estimators will not give the exact same estimates. 
The Kalman filter uses the Minimum Mean Square estimator, for which
\begin{eqnarray}
    \frac{\partial}{\partial \hat{z}} \mathbb{E}\left( \| z - \hat{z} \|^{2} \  \big|  \ \mathbb{M} \right) = 0 \ \ \Rightarrow \ \ \hat{z} = \mathbb{E} \left[z \big| \mathbb{M} \right]
\end{eqnarray}
From now on, we will use the expression $  \hat{z} = \mathbb{E} \left[z \big| \mathbb{M} \right] $ as the estimator. This implies that if we want to estimate the state of the system at time $ k $ we need to find the conditional probability $p\left( z_{k} \big| \mathbb{M}_{k} \right) $ given all measurements $ \mathbb{M}_{k} $.

\subsection{Recursive conditional probability}

The goal of the Kalman algorithm is computing $ \hat{z}_{k}^{k}$, that is, the mean of $ z_{k} $ given all measurements up to time $ k $,
\begin{eqnarray}
    \hat{z}_{k}^{k} = \mathbb{E}\left( z_{k} \big| \mathbb{M} \right) = \int dz_{k} \ z_{k} \ p\left( z_{k} \big| \mathbb{M}_{k} \right)
\end{eqnarray}
For that, we need the probability $p\left( z_{k} \big| \mathbb{M}_{k} \right) $ given all measurements $ \mathbb{M}_{k} $.
We have the relation,
\begin{eqnarray}
    p \left( z_{k} \big| \mathbb{M}_{k} \right) = \frac{p(y_{k} \vert z_{k}) \ p(z_{k} \big| \mathbb{M}_{k-1})}{p(y_{k} \big| \mathbb{M}_{k-1})} \label{iteration}
\end{eqnarray}
This is a consequence of repeated applications of Bayes' rule and joint and conditional probabilities definitions (such as $p(a,b) = p(a\vert b) p(b)$). The main message is that the conditional probability of state $ z_{k} $ given all measurements up to time the present $\mathbb{M}_{k} $ can be expressed in terms of the conditional probability of measurement result $ y_{k} $ at instant $ k $ and conditional probabilities referring to previous time iterations only. In principle, these can all be related to experimentally obtainable quantities. We have the following useful definitions,
\begin{eqnarray}
    p(z_{k} \big| \mathbb{M}_{k-1} ) &=& \int dz_{k-1} \ p(z_{k} \vert z_{k-1}) \ p(z_{k-1} \big| \mathbb{M}_{k-1}) \label{golden_probability1} \\
    p(z_{k} \vert z_{k-1}) &=& \int dw_{k-1} \ \delta(z_{k} - f_{k-1}(z_{k-1}, u_{k-1}, w_{k-1})) \ p(w_{k-1}) \label{golden_probability2}  \\
    p(y_{k} \vert z_{k}) &=& \int dv_{k} \ \delta(y_{k} - h(z_{k-1}, u_{k-1}, v_{k-1})) \ p(v_{k}) \\
    p(y_{k} \big| \mathbb{M}_{k-1}) &=& \int dz_{k} \ p(y_{k} \vert z_{k} ) \ p(z_{k} \big| \mathbb{M}_{k-1})
\end{eqnarray}


\subsection{Kalman estimation}

Consider the linear system 
\begin{eqnarray}
    \Vec{z}_{k} &=& \hat{A} \Vec{z}_{k-1} + \Vec{w}_{k-1} \label{dynamics} \\
    \Vec{y}_{k} &=& \hat{C} \Vec{z}_{k} + \Vec{v}_{k} \label{measuremnent}
\end{eqnarray}
We use the vector and hat symbols to make it explicit that these are multi-dimensional equations. Eq. \eqref{dynamics} is the dynamical law and \eqref{measuremnent} is called the measurement equation. The noise quantities $ \Vec{w}, \Vec{v} $ are zero-mean and have covariance matrices $ \hat{Q}, \hat{R}$, respectively. The initial state has mean $ \Vec{x}_{0} $ and covariance matrix $\hat{\Sigma}_{0}$. We assume $ \hat{A}, \hat{C}, \hat{Q}, \hat{R}, \Vec{x}_{0}, \hat{\Sigma}_{0} $ are known quantities. We seek the variance and mean of $ \Vec{z}_{k} $ conditional on all previous measurements $ \mathbb{M}_{k} $; denote this quantity
\begin{eqnarray}
    \hat{z}_{k}^{k} = \left( \mathrm{mean \ of \ } \Vec{z} \  \mathrm{at \ time } \ k \ \mathrm{given \ all \ previous \ measurements  } \ \mathbb{M}_{k} \right)
\end{eqnarray}
We also adopt the following notation,
\begin{eqnarray}
    \hat{z}_{k}^{k-1} = \left( \mathrm{mean \ of \ } \Vec{z} \  \mathrm{at \ time } \ k \ \mathrm{given \ all \ previous \ measurements  } \ \mathbb{M}_{k-1} \right)
\end{eqnarray}
Note that in this notation, the present time is always shown in the subscript, while the previous measurements are displayed in the superscript. We also have the covariance matrix at time $k$ given $ \mathbb{M}_{k} $,
\begin{eqnarray}
    \Sigma_{k}^{k} = \Sigma_{\mathrm{at \ time \ } k}^{\mathrm{given \ all \ knowlege \ up \ to } \ k} 
\end{eqnarray} 
and the state's covariance matrix at time $k$ given $ \mathbb{M}_{k-1} $,
\begin{eqnarray}
    \Sigma_{k}^{k-1} = \Sigma_{\mathrm{at \ time \ } k}^{\mathrm{given \ all \ knowlege \ up \ to } (k - 1)} 
\end{eqnarray} 

Any Gaussian distribution $ p(\Vec{x})$ has the property,
\begin{eqnarray}
    \ln p(\Vec{x}) &=& -\frac{1}{2} (\Vec{x} - \Vec{\mu})^{T} \hat{\Sigma}^{-1} (\Vec{x} - \Vec{\mu}) + (...\mathrm{other \ terms}...) \\
    &=& -\frac{1}{2} \Vec{x}^{T} \hat{\Sigma}^{-1} \Vec{x} + \Vec{x}^{T} (\hat{\Sigma}^{-1}\Vec{\mu}) + ... \label{log_formula}
\end{eqnarray}
Since $ \Vec{z}_{k}$ is given by a sum of Gaussian distributed random variables, $ p(z_{k} \big| \mathbb{M}_{k} ) $ is Gaussian.
Taking the logarithm of \eqref{iteration},
\begin{eqnarray}
    \ln p \left( \Vec{z}_{k} \big| \mathbb{M}_{k} \right) &=& \ln p(\Vec{y}_{k} \vert \Vec{z}_{k}) + \ln p(\Vec{z}_{k} \big| \mathbb{M}_{k-1}) + (...\mathrm{other \ terms}...) \\ 
    & = & -\frac{1}{2} \left( \Vec{y}_{k} - \hat{C} \Vec{z}_{k} \right)^{T} \hat{R}^{-1} \left( \Vec{y}_{k} - \hat{C} \Vec{z}_{k} \right)  -\frac{1}{2} \left( \Vec{z}_{k} - \hat{z}_{k}^{k-1} \right)^{T} \Sigma_{k}^{k-1} \left( \Vec{z}_{k} - \hat{z}_{k}^{k-1} \right) + ... \\
    &=& \frac{1}{2} \Vec{z}_{k}^{T} \left( \hat{C}^{T} \hat{R}^{-1} C + (\Sigma_{k}^{k-1})^{-1} \right) \Vec{z}_{k} + \Vec{z}_{k}^{T} \left( \hat{C}^{T} \hat{R}^{-1} \Vec{y}_{k} + (\Sigma_{k}^{k-1})^{-1} \hat{z}_{k}^{k-1} \right) + ... \label{log_final}
\end{eqnarray}
Comparing \eqref{log_formula} and \eqref{log_final}, we find
\begin{eqnarray}
    \hat{\Sigma}_{k}^{k} &=& \left( \hat{C}^{T} \hat{R}^{-1} \hat{C} + (\hat{\Sigma}_{k}^{k-1})^{-1} \right)^{-1} \\
    &=& \hat{\Sigma}_{k}^{k-1} - \hat{\mathcal{K}}_{k} \hat{C} \hat{\Sigma}_{k}^{k-1} \label{update_variance1}
\end{eqnarray}
where
\begin{eqnarray}
    \hat{\mathcal{K}}_{k} = \hat{\Sigma}_{k}^{k-1} \hat{C}^{T} \left( \hat{R} \ + \ \hat{C} \  \Sigma_{k}^{k-1}  \ \hat{C}^{T} \right)^{-1}
\end{eqnarray}
is the so-called Kalman matrix. To derive this formula we have used the Woodbury matrix identity, which tells us how a perturbation affects the inverse of a matrix:
\begin{eqnarray}
    \left( A + UCV \right)^{-1} = A^{-1} - A^{-1}U\left( C^{-1} + V A^{-1} U \right)^{-1} V A^{-1}
\end{eqnarray}
Note $ A \Vec{z}_{k-1}$ and $ \vec{w}_{k-1} $ are independent variables and hence,
\begin{eqnarray}
    \Sigma_{k}^{k-1} = \mathrm{Var}\left( A\vec{z}_{k-1} \vert \mathbb{M}_{k-1}   \right) + \mathrm{Var}\left( \vec{z}_{k-1} \vert \mathbb{M}_{k-1}   \right) = A \Sigma_{k-1}^{k-1} A^{T} + Q
\end{eqnarray}
Substituting this relation back in \eqref{update_variance1} we arrive at an update equation for the variance.

Now let us find the update equation for the estimate $\hat{z}_{k}^{k}$. Staring at \eqref{log_formula} and \eqref{log_final}, and using \eqref{update_variance1} we see that,
\begin{eqnarray}
    \hat{z}_{k}^{k} &=& \Sigma_{k}^{k} \left( \hat{C}^{T} \hat{R}^{-1} \Vec{y}_{k} + (\Sigma_{k}^{k-1})^{-1} \hat{z}_{k}^{k-1} \right) \\
    &=& \hat{z}_{k}^{k-1} + \mathcal{K}_{k} \left( \Vec{y}_{k} - \hat{C} \hat{z}_{k}^{k-1} \right)
\end{eqnarray}
In deriving this relation we have used the identity,
\begin{eqnarray}
    (I - (A+B)^{-1}A)B^{-1} = (A+B)^{-1}
\end{eqnarray}
To arrive at the update equation we write,
\begin{eqnarray}
    \hat{z}_{k}^{k-1} = \int d\vec{z}_{k} \ \Vec{z}_{k} \ p(\vec{z}_{k} \vert \mathbb{M}_{k-1})
\end{eqnarray}
We invoke \eqref{golden_probability1} and \eqref{golden_probability2}, and using these relations arrive at
\begin{eqnarray}
    \hat{z}_{k}^{k-1} = \hat{A} \hat{z}_{k-1}^{k-1}
\end{eqnarray}
The Kalman update then becomes,
\begin{eqnarray}
   \hat{z}_{k}^{k} = \hat{A} \hat{z}_{k-1}^{k-1} + \mathcal{K}_{k} \left( \Vec{y}_{k} - \hat{C} \hat{A} \hat{z}_{k-1}^{k-1} \right)
\end{eqnarray}

\subsection{Algorithm}

We summarize the full Kalman algorithm:

\begin{itemize}

    \item[1)] Define covariance of Brownian noise $ Q $;

    \item[2)] Define covariance of the measurement noise $ R $;

    \item[3)] Define mean initial conditions $ \hat{z}_{0}^{0} $ and covariance $ \Sigma_{0}^{0} $;

    \item[4)] Define dynamics and measurement update equations: 
    \begin{eqnarray}
    \vec{z}_{k} &=& A \Vec{z}_{k-1} + \Vec{w}_{k-1} \\
    \vec{y}_{k} &=& C \vec{z}_{k} + \vec{v}_{k}
\end{eqnarray}
    Run simulation for a number $ N $ of steps.

    \item[5)] For $ k $ in $ [1, N]$:

    Compute the Kalman gain matrix
    \begin{eqnarray}
        \hat{\mathcal{K}}_{k} = \hat{\Sigma}_{k}^{k-1} \hat{C}^{T} \left( \hat{R} \ +  \ \hat{C} \  \Sigma_{k}^{k-1}  \ \hat{C}^{T} \right)^{-1}
    \end{eqnarray}
    where
    \begin{eqnarray}
        \Sigma_{k}^{k-1} = A \Sigma_{k-1}^{k-1} A^{T} + Q
    \end{eqnarray}

     Compute the updated covariance matrix,
    \begin{eqnarray}
        \hat{\Sigma}_{k}^{k} = \hat{\Sigma}_{k}^{k-1} - \hat{\mathcal{K}}_{k} \hat{C} \hat{\Sigma}_{k}^{k-1} 
    \end{eqnarray}

     Compute the updated Kalman estimate,
    \begin{eqnarray}
   \hat{z}_{k}^{k} = \hat{A} \hat{z}_{k-1}^{k-1} + \mathcal{K}_{k} \left( \Vec{y}_{k} - \hat{C} \hat{A} \hat{z}_{k-1}^{k-1} \right)
    \end{eqnarray}

    Repeat with $ k \leftarrow k+1$.

\end{itemize}

\section{Simulation}

\subsection{Parameters and equations of motion}

Let us begin by writing the momentum dynamics for the isolated mechanical oscillator representing our tweezed nanoparticle, including stochastic terms due to thermal noise and to measurement backaction:
\begin{equation}\label{eq:isolated_mechanical}
    \dot{p}_z = -\gamma_{th}p_z - m\omega_m^2z + \sqrt{2\gamma_{th}m}f_{p} + \sqrt{4\pi{}K}(\sqrt{\eta}f_{b}(t) + \sqrt{1-\eta}f_{b_\perp}(t))
\end{equation}
$\gamma_{th}$ is the mechanical damping induced by the momentum kicks from the environmental gas. At pressures smaller than $\order{\text{1mbar}}$, it scales linearly with the gas pressure as
\begin{equation} 
    \gamma_{th} = 15.8\frac{R^2p_{gas}}{m_pv_{gas}},
\end{equation}
where R and $m_p$ are the radius and the mass of the nanoparticle, $p_{gas}$ is the gas pressure and $v_{gas}$ is the mean velocity of the gas molecules, obtained from the Maxwell-Boltzmann distribution. $f_i$ (i $\in$ \{p, b, $\text{b}_\perp$\}) are Wiener processes described by
\begin{eqnarray}
    \langle f_{i}(t) \rangle &=& 0 \\
    \langle f_{i}(t)f_{i}(t')\rangle &=& \delta(t-t')(1 + 2\bar{n}\delta_{ip}) \, \ \ i = p, b, b_{\perp}.
\end{eqnarray}
Above, $\bar{n}$ denotes the mean phonon number. Finally, the coefficient $K$ accounts for the strength of the scattering backaction due to measurement of the particle's position with efficiency $\eta$. The order of magnitude of the value used for this constant in the simulations was inspired by experimental results. Given the measured power spectral density of the backaction force $S_{F}^{ba}$, one can obtain K using Eq. (\ref{eq:isolated_mechanical}) via
\begin{equation}
    S_{F}^{ba} = 4\pi{}K = 8.4 \times{} 10^{-41} \text{N}^2\slash\text{Hz}
\end{equation}
where the value in right-hand side was obtained from \cite{magrini2021real}

Next, since this is useful for treating optical and mechanical quadratures equivalently, we turn to dimensionless units by defining $P = \frac{p_z}{p_{z,zpm}}$ and $Z = \frac{z}{z_{zpm}}$, where
\begin{equation}
    z_{zpm} = \sqrt{\frac{\hbar}{2m\omega_m}}
    \qquad
    \text{and}
    \qquad
    p_{z,zpm} = \sqrt{\frac{\hbar{}m\omega_m}{2}}
\end{equation}
In these units, Eq. (\ref{eq:isolated_mechanical}) becomes
\begin{equation}
    \dot{P} = -\gamma_{th}P - \omega_m{}Z + \frac{\sqrt{2\gamma_{th}m}}{p_{z,zpm}}f_{p} + \frac{\sqrt{4\pi{}K}}{p_{z,zpm}}(\sqrt{\eta}f_{b}(t) + \sqrt{1-\eta}f_{b_\perp}(t))
\end{equation}

Now we include the optomechanical equations of motion in our picture. Let $ \mathbf{z} = \mathrm{Tr}(\rho z) $ denote the mean-quadrature vector for the optical and mechanical modes in a state $ \rho $, with 
\begin{eqnarray}
    z = \left[ X,Y,Z,P \right]^{T}
\end{eqnarray}
Here, $ X,Y $ and $ Z,P $ denote the field and mechanical quadrature operators, respectively. In this notation, the second moments of the field and mechanical mode are given by
\begin{eqnarray}
    \mathbf{\Sigma} = \mathrm{Re}\left(   \mathrm{Tr}(\rho ) zz^{T} \right) - \mathbf{z}\mathbf{z}^{T}
\end{eqnarray}
Since we are concerned with Gaussian states, $ \mathbf{z} $ and $ \mathbf{\Sigma} $ are sufficient to completely determine the dynamics.

In the interaction picture, assuming a frame rotating at the frequency of the optical driving input, the equations of motion are given by,
\begin{equation}\label{eq:state-space-model}
\mathbf{\dot{z}}(t) = \mathbf{A}\mathbf{z} + \mathbf{B}{u}(t) + \mathbf{w}(t) + \mathbf{u}_{op}(t)
\end{equation}
where
\begin{equation}
\mathbf{A} = \begin{bmatrix}
-\kappa/2 & \Delta & 0 & 0 \\
-\Delta & -\kappa/2 & -2g & 0 \\
0 & 0 & 0 & \omega_m \\
-2g & 0 & -\omega_m & -\gamma_{th} \\
\end{bmatrix}
,
\qquad
\mathbf{B} = \begin{bmatrix}
0 \\
0 \\
0 \\
1 \\
\end{bmatrix}
,
\qquad
\mathbf{u}_{op} = \begin{bmatrix}
    \sqrt{\kappa}X_{in} \\
    \sqrt{\kappa}Y_{in} \\
    0 \\
    0 \\
\end{bmatrix}
,
\end{equation}
and $ \kappa $ is the cavity linewidth, $ \Delta $ the detuning, $ g $ the coherent scattering interaction frequency, $ \omega_{m} $ the mechanical frequency, $ X_{in}, Y_{in} $ the input optical field quadratures and $\mathbf{w}(t)$ the noise vector given by,
\begin{equation}
    \mathbf{w}(t) = \begin{bmatrix}
        0 \\
        0 \\
        0 \\
        \frac{\sqrt{2\gamma_{th}m}}{p_{z,zpm}}f_p(t) - \frac{\sqrt{4\pi K}}{p_{z,zpm}}(\sqrt{\eta} f_{b}(t)+\sqrt{1-\eta}f_{b_{\perp}}(t)) \\
    \end{bmatrix}
.
\end{equation}

Along with the dynamics of the system, its necessary to define the measurement process, which is expressed as
\begin{equation}
    y(t) = \mathbf{C}\mathbf{z}(t)+y_n(t)
\end{equation}

\noindent with $\mathbf{C}=\begin{bmatrix}
    0 & 0 & 0 & 1
\end{bmatrix}$ and $y_n(t)$ being the measurement noise, this describes continuous measurement of the position of the mechanical oscillator.

The feedback term $u(t)$ is given by
\begin{equation}
    u(t) = -\mathbf{k^T} \mathbf{\hat{z}},
\end{equation}
corresponding to the control law that minimizes the cost function,
\begin{equation}\label{eq:cost-function} 
    J(u) = \lim_{N\xrightarrow[]{}\infty} \Bigg\langle\frac{1}{N} \sum_{k=0}^{N-1}(\mathbf{z_{k}^{T}}\mathbf{Q}\mathbf{z_k} + r u_{k}^2)\Bigg\rangle.
\end{equation}
We shall soon return to the topic of how this optimal control law is determined within our simulations. First, a few details regarding time discretization are going to be addressed.

\subsection{Discrete-time state space}

Although time is continuous, signal processing and control methods require discretization. The continuous system described by Eq. (\ref{eq:state-space-model}) can be discretized via the integral approximation method, based on the assumption that the system input is constant during the sampling period. This assumption is valid when a zero-order hold is used for digital-to-analog conversion and the system is considered to evolve by fixed time-steps $t_k=k\Delta t$, with $\Delta t$ the sampling period. Considering $\mathbf{z}_k = \mathbf{z}(t_k)$, $u_{k} = u(t_k)$, $\mathbf{w}_k = \mathbf{w}(t_k)$ and $\mathbf{u}_{op,k} = \mathbf{u}_{op}(t_k)$, the integral approximation results in the recursive equation,
\begin{equation}
    \begin{split}
    \mathbf{z}_{k+1} & = \exp(\mathbf{A}\Delta t) x_0 + \int_0^{\Delta t}\exp(\mathbf{A}(\Delta t -\tau)) [\mathbf{B}u_0+\mathbf{w}_{k}+\mathbf{u}_{op,k}]\,d\tau\\ 
    & = \mathbf{A_d}\mathbf{z}_k + \mathbf{B_d}u_k + \mathbf{w}_k + \mathbf{u}_{op,k}        
    \end{split}
\end{equation}

\noindent where 
\begin{eqnarray}
    \mathbf{A_d}=\exp(\mathbf{A}\Delta t) \ , \ \ \ \mathbf{B_d}=\int_0^{\Delta t}\exp(\mathbf{A}\tau)\mathbf{B}\,d\tau
\end{eqnarray}
are the discrete-time evolution matrices. Alternatively, $\mathbf{A}_d$ can be determined from the series expansion,
\begin{equation}
    \mathbf{A_d} = \exp(\mathbf{A}\Delta t) = I +\mathbf{A}\Delta t+\frac{\Delta t}{2!}\mathbf{A}^2 + \dots = \sum_{k=0}^\infty \frac{(\Delta t)^k}{k!}\mathbf{A}^k
\end{equation}, 

\noindent and $\mathbf{B_d}$ is given by,

\begin{equation}
    \begin{split}
    \mathbf{B_d}&=\int_0^{\Delta t}\exp(\mathbf{A}\tau)\mathbf{B}\,d\tau =\Bigg(\sum_{k=0}^\infty \frac{(\Delta t)^{k+1}}{(k+1)!}\mathbf{A}^k \Bigg)\mathbf{B}  \\
    &=\Bigg(\sum_{k=1}^\infty \frac{(\Delta t)^{k}}{k!}\mathbf{A}^{k-1} \Bigg)\mathbf{B}=\Bigg(
    -\mathbf{A}^{-1}+\sum_{k=0}^\infty \frac{(\Delta t)^{k}}{k!}\mathbf{A}^{k}\mathbf{A}^{-1} \Bigg)\mathbf{B}\\
    &=\exp(\mathbf{A}\Delta t)[-\exp(-\mathbf{A}\Delta t)\mathbf{A}^{-1}+\mathbf{A}^{-1}]\mathbf{B} \\
    &= (\mathbf{A}_d-\mathbf{I})\mathbf{A}^{-1}\mathbf{B}
    \end{split}
\end{equation}

\subsection{State estimation}

The measurement of the position of the mechanical oscillator at each step is recorded and fed into a Kalman filter in order to estimate the state of the system at all times. The vector $\mathbf{\hat{z}}$ contains the state-estimations and $\mathbf{k^T}$ is the optimal gain calculated by

\begin{equation}
\mathbf{k^T} = (r+\mathbf{B_d^T}\mathbf{S}\mathbf{B_d})^{-1}\mathbf{B_d^T}\mathbf{S}\mathbf{A_d}.   
\end{equation}

Where $\mathbf{S}$ is the solution of the discrete Ricatti equation 

\begin{equation}\label{eq:riccati}
    \mathbf{S} = \mathbf{Q} + \mathbf{A_d}^T\mathbf{S}\mathbf{A_d} - \mathbf{A_d^T}\mathbf{S}\mathbf{B_d}(r+\mathbf{B_d^T}\mathbf{S}\mathbf{B_d})^{-1}\mathbf{B_d^T}\mathbf{S}\mathbf{A_d},
\end{equation}

Eq. (\ref{eq:riccati}) above takes in consideration the discretized time evolution matrices for a state-space representation, $\mathbf{A_d}$ and $\mathbf{B_d}$, the state-cost weighted matrix $\mathbf{Q}$ and the control weight factor $r$.

Simulation of the optomechanical system is carried out by discretizing the dynamics and then implementing a Kalman filter estimator to allow for feedback control. From noisy measurement inputs, the filter outputs estimates for quadrature averages and covariances after each measurement step. It is worth noticing that even though the dynamics of $\mathbf{z}$ are stochastic, the covariance matrix of the system evolves deterministically according to a Lyapunov equation,
\begin{equation}
    \dot{V} = AVA^T + Q 
\end{equation}

\bibliography{main}